# Self - Dual  N = (1,0)  Supergravity  in  Eight  Dimensions with  Reduced  Holonomy  Spin(7)

Hitoshi NISHINO[1]  and  Subhash RAJPOOT[2]

*Department of Physics & Astronomy*
*California State University*
*1250 Bellflower Boulevard*
*Long Beach, CA 90840*

## Abstract

We construct chiral $N = (1,0)$ self-dual supergravity in Euclidean eight-dimensions with reduced holonomy $Spin(7)$, including all the higher-order interactions in a closed form. We first establish the non-chiral $N = (1,1)$ superspace supergravity in eight-dimensions with $SO(8)$ holonomy without self-duality, as the foundation of the formulation. In order to make the whole computation simple, and the generalized self-duality compatible with supersymmetry, we adopt a particular set of superspace constraints similar to the one originally developed in ten-dimensional superspace. The intrinsic properties of octonionic structure constants make local supersymmetry, generalized self-duality condition, and reduced holonomy $Spin(7)$ all consistent with each other.

[1] E-Mail: hnishino@csulb.edu
[2] E-Mail: rajpoot@csulb.edu

## 1. Introduction

One of the most important conclusion out of M-theory [1] is that realistic four-dimensional (4D) theory with chiral fermions can emerge out of 11D supergravity, when the extra 7D dimensions compactify with the reduced holonomy $G_2$ instead of the maximal $SO(7)$ [2][3][4][5]. This has motivated the recent studies of similar manifolds with reduced holonomy, such as 8D with $Spin(7)$ holonomy, or $G_2$, $SU(3)$ and $SU(2)$ holonomies in 7D, 6D and 4D [2][3][4][5]. In particular, the peculiar property of octonions plays an important role in the case of $Spin(7)$ and $G_2$ holonomies [6].

For these manifolds with reduced holonomies, it is well-known that certain self-duality conditions for spin connection play crucial roles [2][3][4]. Moreover, the 7D manifold with $G_2$ holonomy in the compactification of 11D supergravity should be compatible with local supersymmetry, as confirmed by surviving supersymmetry [4][5]. A natural question that follows is how to make local supersymmetry compatible with self-duality conditions with such peculiar reduced holonomies on these manifolds. In other words, the question is how to construct self-dual supergravity theories with reduced holonomies on these manifolds. In the case of self-dual supergravity in 4D, the situation was easier [7][8], because there was no such complication as reduced holonomy. In higher-dimensions, however, reduced holonomies are present, and it becomes more involved to handle them in the presence of self-duality conditions. In this direction, there have been some trials, such as using BRST or topological symmetry as the guiding principle for getting self-dual supergravity in 8D or 7D. However, to our knowledge, there has been no complete self-dual supergravity theory in 8D or 7D with desirable reduced holonomies $Spin(7)$ or $G_2$ [2], including all higher-order terms, based on octonionic structure constants [6][4]. For example, in ref. [9], only the lower-order term level computation has been done, while higher-order non-trivial interactions have been postponed for future studies. The reason seems to be that even though such a formulation looks simple in terms of cohomology at the lowest order, it soon gets extremely involved at cubic interactions or at quartic fermion terms. Moreover, the lagrangians obtained in the topological formulation in [9] are always with gauge-fixings, and hence are not gauge-invariant.

In this paper, we will present the first complete formulation of self-dual supergravity in Euclidean 8D with the reduced $Spin(7)$ holonomy. Even though we are all sure that such a theory must exist in a complete, closed and consistent form, we also know that actual computation would be considerably involved. It turns out to be more than a simple straightforward computation, but it needs more special technology to handle non-trivial higher-order interactions. What we will perform in this paper is to accomplish this difficult mission, namely, to complete the self-dual supergravity in 8D with the reduced holonomy $Spin(7)$, that includes all the higher-order interaction terms, in a self-contained way consistently formulated in superspace.

As a technical tool, we will adopt a very special set of constraints analogous to so-called 'Beta-Function-Favored Constraints' (BFFC), which was originally developed for considerably simplifying complicated $\beta$-function computations for Green-Schwarz superstring in 10D [10]. In 10D, this set of constraints in superspace is so powerful that all the one-loop $\beta$-function computation in 10D is reduced just to a single tadpole diagram to evaluate. We will see how drastically this particular set of constraints simplifies our computation, as well as makes the supersymmetrization of self-duality conditions possible in 8D with the reduced holonomy $Spin(7)$.

## 2. $N = (1,1)$ Supergravity in Euclidean 8D

Before formulating a chiral $N = (1,0)$ superspace with supersymmetric self-duality, we first construct non-chiral $N = (1,1)$ superspace in Euclidean 8D. It is well-known that the set of constraints in a superspace formulation for a given supergravity with fixed field content is not unique. This is due to possible super-Weyl rescalings that can connect different sets of constraints [11]. Our Euclidean 8D case is not an exception here. However, we found that there is a very special set of constraints that are most convenient for imposing supersymmetric self-duality conditions in the next section. This is an 8D analog of the BFFC originally developed in 10D [10]. This set of constraints greatly simplifies the computation, because fermionic components of supercurvatures disappear, and/or no exponential factor with the dilaton field arises in any of constraints [10].

The field content of our $N = (1,1)$ non-chiral supergravity multiplet is $(e_m{}^a, \psi_m{}^{\underline{\alpha}}, C_{mn},$ $A_m, B_m, \chi_{\underline{\alpha}}, \varphi)$. This field content is formally the same as the Minkowskian case [12]. The component fields $A_m$, $B_m$ and $C_{mn}$ have the respective field strengths $F_{mn}$, $G_{mn}$ and $H_{mnr}$. We sometimes call $\chi$ dilatino, while $\varphi$ dilaton. Here we use the *underlined* spinorial indices $\underline{\alpha}, \underline{\beta}, \cdots = 1, 2, \cdots, 16$ in order to distinguish other chiral indices to be used later. These indices are also used for fermionic coordinates as usual in superspace [13], while the indices $m, n, \cdots = 1, 2, \cdots, 8$ are for curved bosonic coordinates, and $a, b, \cdots = 1, 2, \cdots, 8$ are for local Lorentz bosonic coordinates. In the Clifford algebra for Euclidean 8D, we have a symmetric charge conjugation matrix which can be identified with an unit matrix: $C_{\underline{\alpha}\underline{\beta}} = \delta_{\underline{\alpha}\underline{\beta}}$, so that the raising/lowering of spinor indices will not matter, even though we sometimes use their superscripts/subscripts simultaneously, whenever convenient, such as making contractions are more transparent. Relevantly, we have the symmetry $(\gamma_{[n]})_{\underline{\alpha}\underline{\beta}} = +(-1)^{n(n-1)/2}(\gamma_{[n]})_{\underline{\beta}\underline{\alpha}}$ [14], where the symbol $_{[n]}$ implies the totally antisymmetric bosonic indices: $\gamma_{[n]} \equiv \gamma_{a_1 \cdots a_n}$.

The Bianchi identities (BIds) to be satisfied in our superspace are[3]

$$\tfrac{1}{2}\nabla_{[A}T_{BC)}{}^D - \tfrac{1}{2}T_{[AB|}{}^E T_{E|C)}{}^D - \tfrac{1}{4}R_{[AB|f}{}^e(\mathcal{M}_e{}^f)_{|C)}{}^D \equiv 0 ~, \tag{2.1a}$$

$$\tfrac{1}{6}\nabla_{[A}H_{BCD)} - \tfrac{1}{4}T_{[AB|}{}^E H_{E|CD)} + \tfrac{1}{4}F_{[AB}F_{CD)} - \tfrac{1}{4}G_{[AB}G_{CD)} \equiv 0 ~, \tag{2.1b}$$

$$\tfrac{1}{2}\nabla_{[A}F_{BC)} - \tfrac{1}{2}T_{[AB|}{}^D F_{D|C)} \equiv 0 ~, \tag{2.1c}$$

$$\tfrac{1}{2}\nabla_{[A}G_{BC)} - \tfrac{1}{2}T_{[AB|}{}^D G_{D|C)} \equiv 0 ~. \tag{2.1d}$$

After trial and errors, we have found the following BFFC set of constraints in 8D:

$$T_{\underline{\alpha}\underline{\beta}}{}^c = -i(\gamma^c)_{\underline{\alpha}\underline{\beta}} ~, \tag{2.2a}$$

$$H_{\underline{\alpha}\underline{\beta}c} = -\tfrac{i}{2}(\gamma_c)_{\underline{\alpha}\underline{\beta}} ~, \quad F_{\underline{\alpha}\underline{\beta}} = -\tfrac{i}{\sqrt{2}}\delta_{\underline{\alpha}\underline{\beta}} ~, \quad G_{\underline{\alpha}\underline{\beta}} = +\tfrac{i}{\sqrt{2}}(\gamma_9)_{\underline{\alpha}\underline{\beta}} ~, \tag{2.2b}$$

$$T_{\underline{\alpha}\underline{\beta}}{}^{\underline{\gamma}} = -\tfrac{3i}{\sqrt{6}}\big[\,\delta_{\underline{\alpha}\underline{\beta}}\chi^{\underline{\gamma}} + (\gamma^a)_{\underline{\alpha}\underline{\beta}}(\gamma_a\chi)^{\underline{\gamma}} + (\gamma_9)_{\underline{\alpha}\underline{\beta}}(\gamma_9\chi)^{\underline{\gamma}} - \delta_{(\underline{\alpha}}{}^{\underline{\gamma}}\chi_{\underline{\beta})}\,\big] ~, \tag{2.2c}$$

$$T_{\underline{\alpha}b}{}^c = 0 ~, \quad H_{\underline{\alpha}bc} = 0 ~, \quad F_{\underline{\alpha}b} = 0 ~, \quad G_{\underline{\alpha}b} = 0 ~, \tag{2.2d}$$

$$T_{\underline{\alpha}b}{}^{\underline{\gamma}} = -\tfrac{1}{2}(\gamma^{cd})_{\underline{\alpha}}{}^{\underline{\gamma}}H_{bcd} - \tfrac{1}{\sqrt{2}}(\gamma^c)_{\underline{\alpha}}{}^{\underline{\gamma}}F_{bc} - \tfrac{1}{\sqrt{2}}(\gamma_9\gamma^c)_{\underline{\alpha}}{}^{\underline{\gamma}}G_{bc} ~, \tag{2.2e}$$

$$\nabla_{\underline{\alpha}}\varphi = +\tfrac{i}{\sqrt{2}}\chi_{\underline{\alpha}} ~, \tag{2.2f}$$

$$\nabla_{\underline{\alpha}}\chi_{\underline{\beta}} = -\tfrac{1}{\sqrt{2}}(\gamma^c)_{\underline{\alpha}\underline{\beta}}\nabla_c\varphi + \tfrac{1}{6\sqrt{6}}(\gamma^{cde})_{\underline{\alpha}\underline{\beta}}H_{cde} - \tfrac{1}{4\sqrt{3}}(\gamma^{cd})_{\underline{\alpha}\underline{\beta}}F_{cd} - \tfrac{1}{4\sqrt{3}}(\gamma_9\gamma^{cd})_{\underline{\alpha}\underline{\beta}}G_{cd}$$
$$\qquad - \tfrac{3i}{32\sqrt{6}}(\gamma^{cd})_{\underline{\alpha}\underline{\beta}}\chi_{cd} - \tfrac{i}{32\sqrt{6}}(\gamma^{[3]})_{\underline{\alpha}\underline{\beta}}\chi_{[3]} - \tfrac{3i}{32\sqrt{6}}(\gamma_9\gamma^{cd})_{\underline{\alpha}\underline{\beta}}\chi_{9\,cd} - \tfrac{3i}{16\sqrt{6}}(\gamma_9\gamma^c)_{\underline{\alpha}\underline{\beta}}\chi_{9\,c} ~, \tag{2.2g}$$

$$T_{ab}{}^c = +2H_{ab}{}^c ~, \tag{2.2h}$$

$$R_{\underline{\alpha}\underline{\beta}cd} = -\sqrt{2}i\delta_{\underline{\alpha}\underline{\beta}}F_{cd} - \sqrt{2}i(\gamma_9)_{\underline{\alpha}\underline{\beta}}G_{cd} ~, \tag{2.2i}$$

at the mass dimensions $d \leq 1$. Here $\gamma_9 \equiv \gamma_1\gamma_2\cdots\gamma_8$, and $\chi_{[n]} \equiv (\overline{\chi}\gamma_{[n]}\chi)$. There are two important features in these constraints. The first feature is that the fermionic components $H_{\underline{\alpha}bc}$, $F_{\underline{\alpha}b}$, $G_{\underline{\alpha}b}$ are absent. Usually these contain linear dilatino, but in our BFFC these components vanish. This is also related to the second feature that no exponential factor with the dilaton appears anywhere in our constraints. This is very similar to the 10D case [10]. Even though we skip details, we mention that the most frequently-used relationship in these computations is the Fierz identity

$$(\gamma_a)_{(\underline{\alpha}\underline{\beta}}(\gamma^a)_{\underline{\gamma})\underline{\delta}} = \delta_{(\underline{\alpha}\underline{\beta}}\delta_{\underline{\gamma})\underline{\delta}} - (\gamma_9)_{(\underline{\alpha}\underline{\beta}|}(\gamma_9)_{|\underline{\gamma})\underline{\delta}} ~. \tag{2.3}$$

---

[3]As has been well-known [15], the supercurvature BId $\nabla_{[A}R_{BC)d}{}^e + \cdots \equiv 0$ holds automatically, once all of the BIds (2.1) are satisfied.

Some of the BIds at $d \geq 3/2$ are related to superfield equations. The remaining constraints are

$$R_{\underline{\alpha}bcd} = +i(\gamma_b T_{cd})_{\underline{\alpha}} \; , \tag{2.4a}$$

$$\nabla_{\underline{\alpha}} H_{bcd} = -\tfrac{i}{4}(\gamma_{[b} T_{cd]})_{\underline{\alpha}} \; , \qquad \nabla_{\underline{\alpha}} F_{bc} = -\tfrac{i}{\sqrt{2}} T_{bc\underline{\alpha}} \; , \qquad \nabla_{\underline{\alpha}} F_{bc} = +\tfrac{i}{\sqrt{2}}(\gamma_9 T_{bc})_{\underline{\alpha}} \; , \tag{2.4b}$$

$$\begin{aligned}\nabla_{\underline{\gamma}} T_{ab}{}^{\underline{\delta}} =\;& -\tfrac{1}{4}(\gamma^{cd})_{\underline{\gamma}}{}^{\underline{\delta}} R_{cdab} + \tfrac{1}{2}(\gamma^{cd})_{\underline{\gamma}}{}^{\underline{\delta}}(F_{ab} F_{cd} - G_{ab} G_{cd}) + (\gamma_9)_{\underline{\gamma}}{}^{\underline{\delta}} F_{c[a} G_{b]c} \\ & + \tfrac{1}{\sqrt{2}}(\gamma^c)_{\underline{\gamma}}{}^{\underline{\delta}} \nabla_c F_{ab} + \tfrac{1}{\sqrt{2}}(\gamma_9 \gamma^c)_{\underline{\gamma}}{}^{\underline{\delta}} \nabla_c G_{ab} \\ & - \tfrac{3i}{\sqrt{6}} \Big[ T_{ab\underline{\gamma}} \chi^{\underline{\delta}} + (\gamma^c T_{ab})_{\underline{\gamma}}(\gamma_c \chi)^{\underline{\delta}} + (\gamma_9 T_{ab})_{\underline{\gamma}}(\gamma_9 \chi)^{\underline{\delta}} - T_{ab}{}^{\underline{\delta}} \chi_{\underline{\gamma}} - \delta_{\underline{\gamma}}{}^{\underline{\delta}}(\overline{T}_{ab}\chi) \Big] \; , \end{aligned} \tag{2.4c}$$

where expression such as $(\gamma_b T_{cd})_{\underline{\gamma}}$ implies the involvement of the gravitino superfield strength $T_{cd}{}^{\underline{\delta}}$, like $(\gamma_b T_{cd})_{\underline{\gamma}} \equiv (\gamma_b)_{\underline{\gamma}\underline{\delta}} T_{cd}{}^{\underline{\delta}}$. These lead to the gravitino and gravitational, or tensor superfield equations

$$i(\gamma^b T_{ab})_{\underline{\gamma}} + \sqrt{6} i \nabla_a \chi_{\underline{\gamma}} - \tfrac{\sqrt{6}i}{2}(\gamma^{bc}\chi)_{\underline{\gamma}} H_{abc} - \sqrt{3}i(\gamma^b \chi)_{\underline{\gamma}} F_{ab} - \sqrt{3}i(\gamma_9 \gamma^b \chi)_{\underline{\gamma}} G_{ab} \doteq 0 \; , \tag{2.5a}$$

$$R_{ab} - 2(F_{ac} F_b{}^c - G_{ac} G_b{}^c) + 2\sqrt{3} \nabla_a \nabla_b \varphi \doteq 0 \; , \tag{2.5b}$$

$$R_{[ab]} = 2\nabla_c H_{ab}{}^c \doteq -4\sqrt{3} H_{ab}{}^c \nabla_c \varphi - \sqrt{6}i(\overline{T}_{ab}\chi) \; . \tag{2.5c}$$

The symbol $\doteq$ indicates a superfield equation of motion. The first equality in (2.5c) can be obtained from the $T$-BId (2.6a) below at $d = 2$. Relevantly, the superfield equations for dilatino and dilaton are obtained by taking the multiplication of (2.5a) by $\gamma^a$ and the trace of (2.5b), respectively which are skipped here.

There are several remarks in order. First, note the peculiar form of (2.4a), which corresponds to the supersymmetry transformation of the Lorentz connection $\phi_{bcd}$. The point is that the indices $cd$ are on the gravitino superfield strength $T_{cd}{}^{\underline{\beta}}$, which is made possible by the particular choice of the bosonic supertorsion component (2.2h) (Cf. also section 4). Second, similar feature is found in the component $R_{\underline{\alpha}\underline{\beta}cd}$ (2.2i), where the pair of indices $cd$ appears on the superfield strengths $F_{cd}$ and $G_{cd}$. Third, note the particular order of indices $cdab$ on the Riemann supercurvature $R_{cdab}$ in (2.4c). To reach this form, we made use of the identities

$$R_{[abc]}{}^d = -2\nabla_{[a} H_{bc]}{}^d + 4 H_{[ab|}{}^e H_{e|c]}{}^d \; , \tag{2.6a}$$

$$R_{a[bcd]} = -4\nabla_a H_{bcd} + \tfrac{1}{2} F_{[ab} F_{cd]} - \tfrac{1}{2} G_{[ab} G_{cd]} \; , \tag{2.6b}$$

$$\begin{aligned} R_{abcd} - R_{cdab} =\;& -2\nabla_{[a} H_{b]cd} + 4 H_{ab}{}^e H_{ecd} + 4 H_{c[a|}{}^e H_{e|b]d} \\ & - 2 F_{ab} F_{cd} - F_{a[c} F_{d]b} + 2 G_{ab} G_{cd} + G_{a[c} G_{d]b} \; . \end{aligned} \tag{2.6c}$$

Eq. (2.6c) is verified based on (2.6b), while the latter is confirmed by (2.6a) which is nothing but the $T$-BI at $d = 2$. We need the last pair of indices $ab$ free, instead of the first pair $cd$ on $R_{cdab}$ for the following reason. As will be seen, we can *not* impose the self-duality on the first pair of indices of $R_{cdab}$, but only on the last one. This is consistent with the self-duality to be imposed on $T_{cd}{}^{\dot\delta}$.

## 3. Self-Dual $N = (1,0)$ Supergravity in 8D with Reduced Holonomy $Spin(7)$

The non-chiral $N = (1,1)$ supergravity we have established can be a good starting point for the construction of $N = (1,0)$ chiral self-dual supergravity in 8D. Our approach in superspace is analogous to the self-dual supergravity in 4D [7][8], in the sense that the chirality of fermions plays an important role. Our formulation is also a local supersymmetric generalization of our previous globally supersymmetric self-dual Yang-Mills theory in 8D [16], where the peculiar feature of octonionic structure constants [6] was shown to play an important role for self-duality.

Our first prescription is to impose the chirality conditions $\chi^\alpha \stackrel{*}{=} 0$ and $T_{ab}{}^\alpha \stackrel{*}{=} 0$ on fermions, where we use *undotted* (or *dotted*) indices for the positive (or negative) chirality. Needless to say, the ranges of indices are $\alpha, \beta, \cdots = 1, 2, \cdots, 8$ and $\dot\alpha, \dot\beta, \cdots = \dot 1, \dot 2, \cdots, \dot 8$, so that $\underline{\alpha} \equiv (\alpha, \dot\alpha)$, $\underline{\beta} \equiv (\beta, \dot\beta)$, $\cdots$. We also use the symbol $\stackrel{*}{=}$ to elucidate the constraints related to supersymmetric self-duality. This setup is similar to 4D self-dual supergravity [8]. Our next natural conditions are the self-dualities on the curvature supertensor and gravitino superfield strength: $R_{abcd} \stackrel{*}{=} +(1/2) f_{cd}{}^{ef} R_{abef}$, $\overline{T}_{ab}{}^{\dot\gamma} \stackrel{*}{=} +(1/2) f_{ab}{}^{cd} \overline{T}_{cd}{}^{\dot\gamma}$,[4] where $f_{abcd}$ are octonionic structure constant generalized to 8D, satisfying $f_{[4]} = +(1/4!)\epsilon_{[4]}{}^{[4]'} f_{[4]'}$ with $\epsilon_{12\cdots 8} = +1$ [6][4]. However, as in global supersymmetric case [16], these conditions are not enough to guarantee the total consistency. The reason is that some terms produced by applying spinorial derivatives to these conditions may not vanish consistently. The third key point is that we need to impose the extra condition on the *undotted* spinorial components:

$$\nabla_\alpha \equiv \mathcal{P}_{\alpha\beta} \nabla_\beta \equiv (\mathcal{P}\nabla)_\alpha \ , \quad \mathcal{P} \equiv \tfrac{1}{16}(2P + f) \ , \quad f \equiv \tfrac{1}{4!} f^{abcd} \gamma_{abcd} \ , \tag{3.1}$$

where $P$ is the positive chiral projection: $P \equiv (I + \gamma_9)/2$, while $\mathcal{P}$ is the projector of the positive chirality space further into $\mathbf{1}$ in $\mathbf{8} \to \mathbf{7} + \mathbf{1}$ under $SO(8) \to SO(7)$, satisfying $\mathcal{P}^2 \equiv \mathcal{P}$ [4][16]. In other words, even though we do not explicitly write the projector $\mathcal{P}$ for the undotted indices, it should be always understood as an implicit projection from now on. The necessity of this projection is similar to the globally supersymmetric case [16].

---

[4]We use the *bars* for fermions with negative chiralities. This should not be confused with the 'bars' used in the Majorana spinors in the previous section.

Accordingly, we will also see that the field strength $H_{abc}$ and $\nabla_a \varphi$ should be related by octonionic generalized duality, and the field strengths $F_{ab}$ and $G_{ab}$ are proportional to each other, in addition to their own self-dualities.

Eventually, our extra set of constraints needed for supersymmetric self-duality in 8D has been found to be

$$\chi_\alpha \overset{*}{=} 0 ~, \tag{3.2a}$$

$$T_{ab}{}^\gamma \overset{*}{=} 0 ~, \tag{3.2b}$$

$$\overline{T}_{ab}^{(-)\dot\gamma} \equiv N_{ab}{}^{cd}\, \overline{T}_{cd}{}^{\dot\gamma} \overset{*}{=} 0 ~, \quad N_{ab}{}^{cd} \equiv \tfrac{1}{8}(\delta_{[a}{}^c \delta_{b]}{}^d - f_{ab}{}^{cd}) ~, \tag{3.2c}$$

$$\nabla_a \varphi \overset{*}{=} \tfrac{1}{6\sqrt{3}} f_a{}^{bcd} H_{bcd} ~, \tag{3.2d}$$

$$F_{ab}^{(-)} \equiv N_{ab}{}^{cd} F_{cd} \overset{*}{=} 0 ~, \quad G_{ab}^{(-)} \equiv N_{ab}{}^{cd} G_{cd} \overset{*}{=} 0 ~, \tag{3.2e}$$

$$F_{ab}^{(+)} \overset{*}{=} G_{ab}^{(+)} ~, \tag{3.2f}$$

$$R_{abcd}^{(-)} \equiv N_{ab}{}^{ef} R_{abef} \overset{*}{=} 0 ~. \tag{3.2g}$$

Note that we require the self-duality only for the last pair of indices $cd$ on $R_{abcd}$ in (3.2g), but not for the first one. This is due to the existence of torsion component $T_{ab}{}^c$ which makes the $R_{abcd}$ not symmetric under the exchange $ab \leftrightarrow cd$, as (2.6c) shows.

We can confirm the consistency among the constraints in (3.2), by taking their spinorial derivatives, such as $\nabla_\alpha \chi_\beta \overset{?}{=} 0$ for (3.2a). Some remarks are in order for this confirmation, in turn:

(i) $\nabla_{\underline{\alpha}} \chi_\beta \overset{?}{=} 0$: The proof of this equation requires the important identity [6][4]

$$\mathcal{P} \gamma_{(+)}^{ab} \equiv \mathcal{P} P^{abcd} \gamma_{cd} \equiv \gamma_{(+)}^{ab} \mathcal{P} \equiv 0 ~, \quad P_{ab}{}^{cd} \equiv +\tfrac{3}{8}(\delta_{[a}{}^c \delta_{b]}{}^d + \tfrac{1}{3} f_{ab}{}^{cd}) ~, \tag{3.3}$$

which can be confirmed by the use of other more basic identities, such as [6][4]

$$f \gamma_{ab} \equiv -P f_{ab}{}^{cd} \gamma_{cd} - \tfrac{1}{6} P f_{[a|}{}^{[3]} \gamma_{[3]|b]} ~, \tag{3.4a}$$

$$f_{abcd} f^{defg} \equiv -\delta_{[a}{}^e \delta_b{}^f \delta_{c]}{}^g + \tfrac{1}{4} f_{[ab}{}^{[ef} \delta_{c]}{}^{g]} ~. \tag{3.4b}$$

The proof of (3.3) is straightforward but technically involved. Other identities to be used for this sector (i) are [6][4]

$$\gamma_a f \equiv +\tfrac{1}{3} f_a{}^{[3]} \gamma_{[3]} \mathcal{P} ~, \quad f \gamma_a \equiv -\tfrac{1}{3} \mathcal{P} f_a{}^{[3]} \gamma_{[3]} ~, \quad \mathcal{P} f \equiv +14 f \equiv +f \mathcal{P} ~, \tag{3.5a}$$

$$\gamma^{[a|} f \gamma^{|bc]} \mathcal{P} \equiv -24 f^{abcd} \gamma_d \mathcal{P} - 36 \gamma^{abc} \mathcal{P} ~, \tag{3.5b}$$

$$\gamma_{[a} f \gamma_{b]} \equiv +2 N f_{ab}{}^{cd} \gamma_{cd} ~, \quad f_{abcd} N^{cdef} \equiv -6 N_{ab}{}^{ef} ~. \tag{3.5c}$$

(ii) $\nabla_{\underline{\alpha}} T_{ab}{}^\delta \overset{?}{=} 0$: The proof of this equation requires the important identities

$$\mathcal{P}\gamma_{ab}\mathcal{P} \equiv 0 \ , \tag{3.6a}$$

$$(\gamma^c \overline{T}_{ab})_\gamma (\gamma_c \overline{\chi})^\delta \equiv \delta_\gamma{}^\delta (\overline{T}_{ab}\overline{\chi}) - \tfrac{1}{4}(\gamma^{de})_\gamma{}^\delta (\overline{T}_{ab}\gamma_{de}\overline{\chi}) \ . \tag{3.6b}$$

The identity (3.6a) can be proven with the aid of more basic ones, such as

$$\begin{aligned} \{\mathcal{P}, \gamma_{ab}^{(-)}\} &\equiv +P\gamma_{ab}^{(-)} \ , \\ \{f, \gamma_{ab}\} &\equiv -2P f_{ab}{}^{cd}\gamma_{cd} \ . \end{aligned} \tag{3.7}$$

(iii) $\nabla_{\underline{\alpha}} \overline{T}_{ab}^{(-)\dot\delta} \overset{?}{=} 0$: This equation is easy to confirm by the vanishing of the r.h.s. Note, however, the important fact that the indices $ab$ on the curvature supertensor $R_{cdab}$ satisfy the self-duality condition (3.2g). This enables us to show the vanishing of the r.h.s. after taking the spinorial derivative.

(iv) $\nabla_{\underline{\alpha}}[\nabla_a\varphi - (6\sqrt{6})^{-1} f_a{}^{[3]} H_{[3]}] \overset{?}{=} 0$: This equation is rather straightforward, and we can easily show that all the terms in the r.h.s. vanish, upon using the constraints in (3.2).

(v) $\nabla_{\underline{\alpha}} F_{ab}^{(-)} \overset{?}{=} 0$ and $\nabla_{\underline{\alpha}} G_{ab}^{(-)} \overset{?}{=} 0$: These are easy to prove, upon using the self-duality (3.2c).

(vi) $\nabla_{\underline{\alpha}}(F_{ab} - G_{ab}) \overset{?}{=} 0$: This equation is also easy to satisfy with no nontrivial identity needed.

(vii) $\nabla_{\underline{\alpha}} R_{bcde}^{(-)} \overset{?}{=} 0$: The proof of this equation needs the supercurvature BId at $d = 5/2$, starting with $\nabla_{\underline{\alpha}} R_{bcde} + \cdots \equiv 0$. Interestingly enough, when the first term is expressed in terms of the remaining terms, after the projection of $de$ into the anti-self-dual components, all the terms vanish upon using the constraints in (3.2). This concludes the confirmation of all the spinorial derivatives of the constraints in (3.2).

As the main result of this paper, we collect below our superspace constraints for $N = (1,0)$ self-dual supergravity with the reduced holonomy $Spin(7)$:[5]

$$T_{\alpha\dot\beta}{}^c = -i(\gamma^c)_{\alpha\dot\beta} \ , \quad T_{\dot\alpha\beta}{}^c = -i(\gamma^c)_{\dot\alpha\beta} \ , \tag{3.8a}$$

$$T_{\alpha\beta}{}^\gamma \overset{*}{=} T_{\alpha\beta}{}^{\dot\gamma} \overset{*}{=} T_{\dot\alpha\dot\beta}{}^\gamma \overset{*}{=} T_{\dot\alpha\dot\beta}{}^{\dot\gamma} \overset{*}{=} 0 \ , \tag{3.8b}$$

$$H_{\alpha\dot\beta c} = -\tfrac{i}{2}(\gamma_c)_{\alpha\dot\beta} \ , \tag{3.8c}$$

$$F_{\alpha\beta} \overset{*}{=} -\tfrac{i}{\sqrt{2}}\delta_{\alpha\beta} \ , \quad F_{\dot\alpha\dot\beta} \overset{*}{=} -\tfrac{i}{\sqrt{2}}\delta_{\dot\alpha\dot\beta} \ , \quad F_{\alpha\dot\beta} \overset{*}{=} 0 \ , \tag{3.8d}$$

---

[5] Note that the *bar*-symbol in expressions like $(\overline{\chi}\gamma_{cd}\overline{\chi}) \equiv \overline{\chi}_{\dot\alpha}(\gamma_{cd})_{\dot\alpha\dot\beta}\overline{\chi}_{\dot\beta}$ symbolizes the *dottedness*, different from the case for Majorana spinors in the last section.

$$G_{\alpha\beta} \stackrel{*}{=} +\tfrac{i}{\sqrt{2}}\delta_{\alpha\beta} \ , \qquad G_{\dot\alpha\dot\beta} \stackrel{*}{=} -\tfrac{i}{\sqrt{2}}\delta_{\dot\alpha\dot\beta} \ , \qquad G_{\alpha\dot\beta} \stackrel{*}{=} 0 \ , \tag{3.8e}$$

$$T_{\alpha\beta}{}^{\gamma} \stackrel{*}{=} -\tfrac{3i}{\sqrt{6}}\Big[(\gamma^{d})_{\alpha\dot\beta}(\gamma_{d}\overline\chi)^{\gamma} - \delta_{\alpha}{}^{\gamma}\overline\chi_{\dot\beta}\Big] \stackrel{*}{=} +T_{\beta\dot\alpha}{}^{\gamma} \ , \tag{3.8f}$$

$$T_{\dot\alpha\dot\beta}{}^{\dot\gamma} \stackrel{*}{=} -\tfrac{3i}{\sqrt{6}}\Big[2\delta_{\dot\alpha\dot\beta}\overline\chi^{\dot\gamma} - \delta_{(\dot\alpha}{}^{\dot\gamma}\overline\chi_{\dot\beta)}\Big] \ , \tag{3.8g}$$

$$T_{ab}{}^{\gamma} \stackrel{*}{=} -\tfrac{1}{2}(\gamma^{cd})_{\alpha}{}^{\gamma}H_{bcd} \ , \qquad T_{\dot a b}{}^{\dot\gamma} \stackrel{*}{=} -\tfrac{1}{2}(\gamma^{cd})_{\dot\alpha}{}^{\dot\gamma}H_{bcd} \ , \tag{3.8h}$$

$$T_{ab}{}^{\dot\gamma} \stackrel{*}{=} -\sqrt{2}(\gamma^{c})_{\alpha}{}^{\dot\gamma}F_{bc} \ , \tag{3.8i}$$

$$\overline\nabla_{\dot\alpha}\varphi \stackrel{*}{=} \tfrac{i}{\sqrt{2}}\overline\chi_{\dot\alpha} \ , \qquad \nabla_{\alpha}\varphi \stackrel{*}{=} 0 \ , \tag{3.8j}$$

$$\overline\nabla_{\dot\alpha}\overline\chi_{\dot\beta} \stackrel{*}{=} -\tfrac{3i}{16\sqrt{6}}(\gamma^{cd})_{\dot\alpha\dot\beta}(\overline\chi\gamma_{cd}\overline\chi) \ , \tag{3.8k}$$

$$R_{\alpha\beta cd} \stackrel{*}{=} -2\sqrt{2}\delta_{\alpha\beta}F_{cd} \ , \qquad R_{\dot\alpha\dot\beta cd} \stackrel{*}{=} R_{\alpha\dot\beta cd} \stackrel{*}{=} 0 \ , \tag{3.8$\ell$}$$

$$R_{\alpha bcd} \stackrel{*}{=} +i(\gamma_{b}\overline T_{cd})_{\alpha} \ , \qquad R_{\dot\alpha bcd} \stackrel{*}{=} 0 \ , \qquad R^{(-)}_{abcd} \stackrel{*}{=} 0 \ , \tag{3.8m}$$

$$\nabla_{\alpha}H_{bcd} \stackrel{*}{=} -\tfrac{i}{4}(\gamma_{[b}T_{cd]})_{\alpha} \ , \qquad \overline\nabla_{\dot\alpha}H_{bcd} \stackrel{*}{=} 0 \ , \tag{3.8n}$$

$$\overline\nabla_{\dot\alpha}F_{bc} \stackrel{*}{=} -\tfrac{i}{\sqrt{2}}\overline T_{bc\dot\alpha} \ , \qquad \nabla_{\alpha}F_{bc} \stackrel{*}{=} 0 \ , \tag{3.8o}$$

$$\nabla_{\gamma}\overline T_{ab}{}^{\dot\delta} \stackrel{*}{=} +\sqrt{2}(\gamma^{c})_{\gamma}{}^{\dot\delta}\nabla_{c}F_{ab} \ , \tag{3.8p}$$

$$\overline\nabla_{\dot\gamma}\overline T_{ab}{}^{\dot\delta} \stackrel{*}{=} -\tfrac{1}{4}(\gamma^{cd})_{\dot\gamma}{}^{\dot\delta}R_{abcd} - \tfrac{3i}{\sqrt{6}}\Big[2\overline T_{ab\dot\gamma}\overline\chi^{\dot\delta} - \overline T_{ab}{}^{\dot\delta}\overline\chi_{\dot\gamma} - \delta_{\dot\gamma}{}^{\dot\delta}(\overline T_{ab}\overline\chi)\Big] \ . \tag{3.8q}$$

As has been stated before, all the *undotted* spinorial components are projected into the $\mathcal{P}$-subspace, *e.g.*, $T_{\alpha\dot\beta}{}^{c} = -i(\gamma^{c})_{\alpha\dot\beta} = -i(\mathcal{P}\gamma^{c})_{\alpha\dot\beta}$ which we do not write explicitly. Needless to say, the component supersymmetry transformation rule can be easily constructed from these constraints, following the general prescription in p. 323 of [13].

Finally, our gravitino, graviton, and the antisymmetric tensor superfield equations turn out to be very simple in our system

$$i(\gamma^{b}\overline T_{ab})_{\gamma} - 2\sqrt{3}i(\gamma^{b}\overline\chi)_{\gamma}F_{ab} \stackrel{*}{=} 0 \ , \tag{3.9a}$$

$$i\nabla_{a}\overline\chi_{\dot\gamma} - \tfrac{i}{2}(\gamma^{bc}\overline\chi)_{\dot\gamma}H_{abc} \stackrel{*}{=} 0 \ , \tag{3.9b}$$

$$R_{ab} \stackrel{*}{=} -2\sqrt{3}\nabla_{a}\nabla_{b}\varphi \ , \tag{3.9c}$$

$$R_{[ab]} = +2\nabla_{c}H_{ab}{}^{c} \stackrel{*}{=} -4\sqrt{3}H_{ab}{}^{c}\nabla_{c}\varphi - \sqrt{6}i(\overline T_{ab}\overline\chi) \ . \tag{3.9d}$$

We use the symbol $\stackrel{*}{=}$ also for superfield equations here, to stress the usage of self-duality conditions in (3.2). By taking the $\gamma^{a}$-multiplication by (3.9a) or (3.9b), and taking the $ab$-trace of (3.9c), we get respectively the dilatino and dilaton superfield equations which are similar to the 10D case of BFFC [10].

## 4. Reduced Holonomy $Spin(7)$

As has been mentioned, (3.2g) implies the consistency of our system with reduced $Spin(7)$ holonomy [2][4][5]. This is because the supersymmetry transformation of the Lorentz spinor connection $\phi_{bcd}$ is given [13] by (3.8m) as

$$\delta_Q \phi_{bcd} = -i(\epsilon\gamma_b \overline{T}_{cd}) \ , \tag{4.1}$$

which is consistent with the self-duality of the last two indices on $\phi_{bcd}$:

$$\phi_{bcd} \overset{*}{=} +\tfrac{1}{2} f_{cd}{}^{ef} \phi_{bef} \ , \tag{4.2}$$

and is based on the self-duality $\overline{T}_{cd}{}^{\dot\delta} = +(1/2)f_{cd}{}^{ef}\overline{T}_{ef}{}^{\dot\delta}$. This also means that the original $SO(8)$ holonomy has been now reduced to $Spin(7)$, and the indices $cd$ on $\phi_{bcd}$ are in the **21** of $Spin(7)$ reduced from the **28** of $SO(8)$. The important ingredient here is that our BFFC have been compatible with such a requirement. If we choose a different set of constraints away from the BFFC, then there generally arise additional terms in (4.1), which result in

$$\delta'_Q \phi_{bcd} = -i(\epsilon\gamma_b \overline{T}_{cd}) + \alpha(\epsilon\gamma_{[c}\overline{T}_{d]b}) \qquad (\alpha\neq 0) \ , \tag{4.3}$$

which are not compatible with the self-duality of the original $\phi_{bcd}$, because of the index structure of the last term. To avoid this obstruction, we have made the bosonic torsion component $T_{ab}{}^c$ (2.2h) non-vanishing, in order to absorb such unwanted terms.

The gravitational superfield equation (3.9c) is also consistent with the self-duality of the Riemann supercurvature. This is easily seen by using the identity (2.6b), as

$$\begin{aligned} R_{ac} &= \delta^{bd} R_{abcd} \overset{*}{=} +\tfrac{1}{2} f_c{}^{def} R_{adef} = +\tfrac{1}{12} f_c{}^{def} R_{a[def]} \\ &= +\tfrac{1}{12} f_c{}^{def}\Big(-4\nabla_a H_{def} + \tfrac{1}{2} F_{[ad} F_{ef]} - \tfrac{1}{2} G_{[ad} G_{ef]}\Big) \\ &\overset{*}{=} -2\sqrt{3}\nabla_a\Big(\tfrac{1}{6\sqrt{3}} f_c{}^{def} H_{def}\Big) \overset{*}{=} -2\sqrt{3}\nabla_a\nabla_c\varphi \ . \end{aligned} \tag{4.4}$$

This is nothing but the torsion-full generalization of the usual Ricci flatness derived from the self-duality of a torsion-less Riemann tensor [4].

As careful readers may have noticed, the counting of physical degrees of freedom in this system needs special care. First of all, the self-dual graviton $e_m{}^a$ has $(3\times 4)/2 - 1 = 5$ on-shell degrees of freedom, where the number 3 is from the self-duality condition that halves the original transversal components $8-2 = 6$ into 3. Similarly, for the self-dual chiral gravitino $\overline{\psi}_m{}^{\dot\alpha}$, the degrees of freedom counts like $3\times 4 = 12$, where 3 is from the self-duality, while 4 is from the chirality. In the case of the tensor $C_{mn}$ and the dilaton $\varphi$, the former has originally $(6 \times 5)/2 = 15$ degrees of freedom, while $\varphi$ has one. After imposing the condition (3.2d), the total $15+1$ will be halved to $8$. For the vectors $A_m$ and $B_m$, they have originally $6+6$ transversal degrees of freedom, which are also halved to be $3+3$ by the duality conditions (3.2e). However, we have additional relationship (3.2f) between them reducing $3+3$ further down to $3$ degrees of freedom. The dilatino $\overline{\chi}^{\dot{\alpha}}$ has simply $4$ on-shell degrees of freedom as a chiral spinor in 8D. In total, we have $5+8+3 = 16$ for the bosons, and $12+4 = 16$ for the fermions, *i.e.,* the balance of on-shell degrees of freedom is $16+16$. Thus we see a considerable reduction in the degrees of freedom from the original $48+48$ before imposing any supersymmetric self-duality conditions (3.2).

Note that the success of our formulation is based on the elaborate combination of the peculiar feature of the octonionic structure constant $f_{abcd}$ leading to generalized self-duality, the fermionic chirality projection, and the usage of BFFC constraints, all closely related to each other consistently in superspace.

## 5. Concluding Remarks

In this paper, we have established self-dual chiral $N = (1,0)$ supergravity in Euclidean 8D, with the reduced holonomy $Spin(7)$. To our knowledge, this is the first *complete* version of such a formulation for local supersymmetry in 8D, including all the higher-order terms, based on the octonionic structure constants $f_{abcd}$. After imposing the supersymmetric generalized self-duality conditions in (3.2), we see that the original $48+48$ on-shell degrees of freedom of the whole multiplet are reduced to $16+16$.

We see that the special features of octonions make it possible to formulate such generalized self-duality conditions compatible with local supersymmetry in 8D. This has been realized by adopting a very particular set BFFC in 10D. The BFFC not only drastically simplify the computation, but also make the generalized self-duality conditions compatible with supergravity. Our result here is made possible by special combinations of superspace technology in the past, such as BFFC and self-duality supergravity theories in 4D, and most importantly, the peculiar feature of the octonionic structure constants $f_{abcd}$ associated with the projector $\mathcal{P}$. We see that the usual concept of self-duality is generalized in terms of $f_{abcd}$ in 8D, as in the globally supersymmetric case [6][4][16].

It has been expected that a formulation for self-dual supergravity in 8D should be highly 'geometrical', and thus mathematically beautiful. However, we have found this statement only 'partially' true. It is true in the sense that our superspace formulation gave a very clear description of such self-duality in 8D supergravity *via* geometrical BIds. It is also true in the sense that the very special choice of BFFC drastically simplified the whole computation, such as disappearance of dilaton exponents, or vanishing fermionic components $F_{\underline{\alpha}b}$, $G_{\underline{\alpha}b}$ and $H_{\underline{\alpha}bc}$. Nevertheless, we also point out that the actual supergravity is more involved than so-called Chern-Simons (CS) type theory formulated in terms of cohomology and superalgebra [17], or BRST topological formulation [9]. The reason is that in contrast to CS-type formulation, 'Poincaré' supergravity is based on actual diffeomorphisms for translations. A typical comparison can be done with so-called CS supergravity [17], where local 'supersymmetries' are realized in terms of superalgebra. The difference in these theories from our 'Poincaré' supergravity is that the commutators of two supersymmetries in CS theory [17] do not yield diffeomorphism as the translation operation on all the fields.[6] This diffeomorphism associated with Clifford algebra makes the whole computation considerably non-trivial. Additionally, it is this non-triviality that has necessitated the case of the particular set of BFFC.

Our work here can be viewed as the locally supersymmetric generalization of our previous work [16] on global supersymmetry in 8D [18]. As it is clear for readers who followed technical details, the peculiar features of octonionic structure constant used in this paper are parallel to global supersymmetry [18][16]. Even though global self-dual supersymmetry was easier in component language [16], the success of our formulation in this paper is due to the usage of superspace language. This is because a parallel computation in component would generate messy higher-order terms, such as quartic fermion terms requiring considerable effort to manage in a closed form. Additionally, even though there has been superspace formulation for global supersymmetries for generalized self-dualities in $D > 4$ [19], our result here is compatible with local supersymmetry thanks to the particular set of BFFC. It is the superspace formulation with the particular BFFC that played a decisive role for the simplified consistency between octonionic structure constant, supersymmetry, fermionic chirality and generalized self-duality in 8D.

There are similarities and differences between our self-dual supergravity in 8D and self-dual supergravity in 4D [7][8]. The most important similarity is the role of chiral spinors playing a crucial role, closely related to the elimination of the anti-self-dual components. The most important difference is the introduction of the reduced holonomy $Spin(7)$ governed by octonionic structure constant $f_{abcd}$. Even though the reduction of holonomy from $SO(2,2)$ into $SU(1,1)$ also exists in 4D self-dual supergravity correlated with chirality projection [7][8], our 8D case differs due to the crucial role played by $f_{abcd}$.

We have dealt in this paper with the holonomy $Spin(7)$ reduced from the maximal one

---

[6]For example, the commutator of two supersymmetries on the tensor field $b^{abcde}$ in 11D CS supergravity in [17] does not induce the usual diffeomorphism on this field.

$SO(8)$ in 8D. We stress this important feature of supergravity is in a sense analogous to the recent development in supergravity with no manifest Lorentz covariances in dimensions $D \geq 12$ [20]. It seems that this feature of reduced or non-manifest holonomy has become more and more common in higher-dimensional supergravity theories in $D \geq 4$.

Our formulation has been performed only in compact Euclidean 8D, that has the closest relationship with compactifications. However, we also mention an alternative possibility of formulating with non-definite signature, such as $D = 4+4$. This is because there is a different set of octonions in 7D with signature $D = 4+3$ [21]. Such a theory may have closer link with integrable systems in lower-dimensions, as the self-dual theories in $D = 2+2$ that have links with integrable models in $D \leq 3$ [22][7][8].

Our theory in this paper can also serve as an underlying theory of all the possible lower-dimensional self-dual supergravity theories in $D \leq 7$. This is because once superspace formulation has been established in 8D, we can apply the dimensional reduction technique developed for dealing with the backgrounds for Green-Schwarz superstring in [23]. Such dimensional reductions in superspace are supposed to generate lower-dimensional (self-dual) supergravities in $D \leq 7$, similarly to the component case from 8D into 7D for global supersymmetric self-dual theories demonstrated in our recent paper [16].

*Note added:* After this paper had been completed, we came across a new paper [24] where self-dual supergravity in 8D based on BRST symmetry was presented with lower-order terms after gauge-fixings, in essential agreement with ours.